\documentclass[aps,twocolumn,prd,showpacs,amsmath,amssymb,superscriptaddress,sort&compress,preprintnumbers]{revtex4-1}
\usepackage{graphicx,amsmath,amssymb,hyperref,natbib,slashed}
\usepackage[dvipsnames]{xcolor}
\usepackage{xspace}

\usepackage[utf8]{inputenc}
\usepackage[normalem]{ulem}

\newcommand{\eVdist}{\kern-0.06em}

\newcommand{\be}{\begin{align}}
\newcommand{\ee}{\end{align}}
\newcommand{\bea}{\begin{eqnarray}}
\newcommand{\eea}{\end{eqnarray}}

\newcommand{\nc}{N_{c_D}}                
\newcommand{\nf}{N_{f_D}}                
\newcommand{\pid}{{\pi_D}}                
\newcommand{\rhod}{{\rho_D}}                
\newcommand{\mpi}{{m_{\pi_D}}}             
\newcommand{\mrho}{{m_{\rho_D}}}             
\newcommand{\fpi}{f_{\pi_D}}             

\begin{document}

\title{Dark matter relic density in strongly interacting dark sectors with light vector mesons}

\newcommand{\AddrFNAL}{%
Particle Theory Department, Fermilab, Batavia, IL 60510, USA
}
\newcommand{\AddrKIT}{%
Institute for Theoretical Particle Physics (TTP), Karlsruhe Institute of Technology (KIT), 76128 Karlsruhe, Germany}

\newcommand{\AddrGraz}{%
Institute of Physics, NAWI Graz, University of Graz, Universit\"atsplatz 5, 8010 Graz, Austria}

 \author{Elias Bernreuther}
 \email{ebernreu@fnal.gov}
 \affiliation{\AddrFNAL}
 
  \author{Nicoline Hemme}
 \email{nicoline.hemme@kit.edu}
 \affiliation{\AddrKIT}
 
 \author{Felix Kahlhoefer}
 \email{felix.kahlhoefer@kit.edu}
 \affiliation{\AddrKIT}
 
  \author{Suchita Kulkarni}
 \email{suchita.kulkarni@uni-graz.at}
 \affiliation{\AddrGraz}

\begin{abstract}
Stable dark matter particles may arise as pseudo-Goldstone bosons from the confinement of dark quarks interacting via a non-Abelian gauge force. Their relic abundance is determined not by annihilations into visible particles but by dark pion number-changing processes within the dark sector, such as $3 \pid \to 2 \pid$. However, if the dark vector mesons $\rhod$ are light enough for $3 \pid \to \pid \rhod$ annihilations to be kinematically allowed, this process dominates and significantly delays freeze-out. As a result, the preferred dark matter mass scale increases and bounds from the Bullet Cluster can be evaded.
\end{abstract}

\preprint{FERMILAB-PUB-23-744-T, TTP23-057, P3H-23-096}

\maketitle

\paragraph*{Introduction.---}%
An attractive alternative to the paradigm of weakly interacting massive particles is the idea that dark matter (DM) is part of a strongly interacting dark sector~\cite{Kribs:2016cew}. At high energies, such a dark sector can be described in terms of dark quarks interacting via the dark gluons of a non-Abelian extension of the Standard Model (SM) gauge group. At low energies, on the other hand, the dark sector confines, and the dark quarks and gluons are bound in dark mesons and dark baryons. In analogy to SM quantum chromodynamics, the pseudoscalar mesons, called dark pions, are expected to be the lightest dark sector state, because they are the pseudo-Goldstone bosons of chiral symmetry breaking. Indeed, if the dark quark masses are sufficiently small, the dark pions can be significantly lighter than the confinement scale. In contrast to the SM, however, the dark pions may be stable, due to either a $U(1)$ or a parity symmetry~\cite{Bai:2010qg,Buckley:2012ky,Cline:2013zca} (although the latter may be broken through gravitational effects~\cite{Dondi:2019olm}), making them attractive DM candidates~\cite{Hochberg:2014kqa, Hochberg:2015vrg,Bernreuther:2019pfb}.

Strongly interacting dark sectors are an attractive target for collider searches due to their striking signature: If a pair of dark quarks is produced in a hard process, it will generate a shower of dark hadrons, some of which may decay into SM particles, while others evade detection~\cite{Strassler:2006im,Englert:2016knz,Pierce:2017taw,Kribs:2018ilo,Cheng:2019yai,Beauchesne:2019ato,Butterworth:2021jto,Cheng:2021kjg,Bernreuther:2022jlj,Albouy:2022cin}. The results are one or more semi-visible~\cite{Cohen:2015toa,Beauchesne:2017yhh} or emerging~\cite{Schwaller:2015gea} jets. These signatures have been explored in a bottom-up way by varying phenomenological parameters such as the masses of dark mesons and the fraction of invisible particles~\cite{Cohen:2017pzm,Cohen:2020afv,Bernreuther:2020vhm,Knapen:2021eip}, and corresponding searches have been carried out by ATLAS~\cite{ATLAS:2023swa} and CMS~\cite{CMS:2021dzg}.  
In order to make a connection to the DM puzzle, however, it becomes necessary to explore the cosmological history of strongly interacting dark sectors, and to understand whether the dark pions can be produced in the right amount to explain observations of the Cosmic Microwave Background~\cite{Bernreuther:2019pfb}.

A central result in this context is the so-called SIMP (for strongly interacting massive particle) mechanism~\cite{Hochberg:2014dra,Hochberg:2014kqa}: dark pions may participate in number-changing processes, such as $3 \pid \to 2 \pid$ via the Wess-Zumino-Witten anomaly. At low temperatures, these processes lead to the conversion of rest mass into kinetic energy and hence a depletion of the dark sector. The freeze-out temperature, when number-changing processes become inefficient, then determines the relic abundance of dark pions. Comparison to observations leads to a strict upper bound of $m_\pid \lesssim 100 \, \mathrm{MeV}$ in order to avoid an overabundance of DM~\cite{Hochberg:2014kqa, Hansen:2015yaa,Choi:2017mkk,Braat:2023fhn}. This bound is in tension with a lower bound on the dark pion mass obtained from observations of the Bullet Cluster, which constrains the DM self-interaction cross section~\cite{Randall:2008ppe,Robertson:2016xjh,Wittman:2017gxn}.

In this letter we point out an important modification of the SIMP mechanism: In addition to number-changing processes involving only dark pions, there may also be processes involving heavier dark sector states, in particular the vector mesons, called dark rho mesons~\cite{Choi:2018iit,Berlin:2018tvf,Beauchesne:2018myj}. Indeed, if the dark quarks have masses comparable to the confinement scale, it is possible to have $m_\rhod < 2 m_\pid$. This scenario is particularly interesting for collider experiments, since the dark rho meson in this case cannot decay into dark pions and must instead decay into SM particles (e.g. via kinetic mixing with the SM photon). The lifetime of the dark rho mesons can be quite long, such that their decays lead to displaced vertices in the detector~\cite{Bernreuther:2020xus,Bernreuther:2022jlj}.  

We show that, if the process $3 \pid \to \pid \rhod$ is kinematically allowed at low temperatures, it will typically dominate the rate of number-changing processes. This is due to a favourable velocity dependence (the process proceeds via $s$-wave, whereas the process involving only pions proceeds via $d$-wave) and a resonant enhancement if the internal dark pions can be nearly on-shell. We calculate the relevant cross sections, discuss how thermal effects determine the width of the resonance, and solve the resulting Boltzmann equation. Our central result is that the presence of dark rho mesons relaxes the cosmological bound on the dark pion mass. It thus becomes possible to realize the SIMP mechanism for heavier dark pions and smaller couplings, thereby evading the Bullet Cluster constraint.

\begin{figure*}[t!]
\centering
\includegraphics[width=\textwidth]{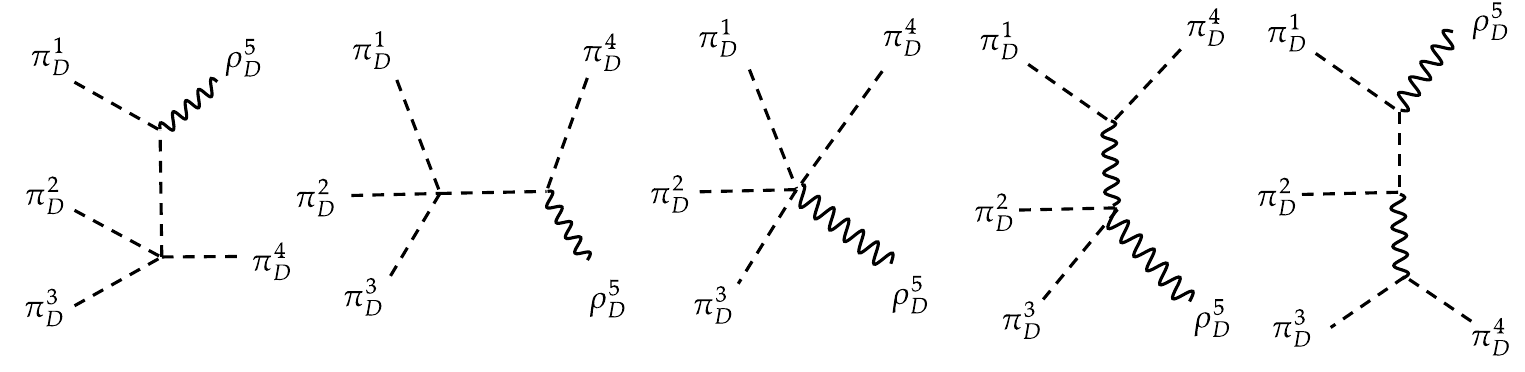}
\caption{Representative Feynman diagrams for the processes that contribute to the $3\pid \to \pid\rhod$ cross section. Additional diagrams are obtained from the first, third, fourth and fifth diagram through permutations of the pion states.}
\label{fig:3pi_to_pi_rho}
\end{figure*}

The remainder of this letter is structured as follows. We first introduce the strongly interacting dark sector model and discuss the spectrum of dark mesons expected from non-perturbative calculations. We then present the relevant Boltzmann equations and the required reaction rates and finally calculate the dark pion relic density and the Bullet Cluster constraint.

\smallskip

\paragraph*{Model details.---}%

We start from an $SU(\nc)$ gauge group, with $\nf$ mass-degenerate Dirac fermions $q_{D}$ in its fundamental representation. The high-energy Lagrangian for such a theory is given by 
\begin{align}
\mathcal{L}_\text{UV} = -\frac{1}{4}G_{D\mu\nu}^aG^{\mu\nu, a}_D + \bar{q}_D(i\gamma^\mu D_\mu - M_{q_D})q_D \, ,
\end{align}
where $M_{q_D}$ is the mass of the dark quarks, $G_{D}^{\mu\nu}$ denotes the dark gluon field strength tensor and $D_\mu$ is the gauge covariant derivative. While we leave $\nc, \nf$ as free parameters, it is important to note that chiral symmetry breaking only takes place for $\nf < 3\nc$.  

At low energies, chiral symmetry breaking leads to confinement. The dynamics of the resulting pseudo-Nambu-Goldstone Bosons, called dark pions $\pid$, are described by a chiral Lagrangian. The kinetic term and the mass term are contained in
\begin{align}
\label{eq:chiral_lagrangian}
\mathcal{L}_\text{Ch} = \tfrac{\fpi^2}{4}\mathrm{Tr}\left(\partial_\mu U\partial^\mu U^\dagger\right) + \left[\tfrac{\mu_D^3}{2}\mathrm{Tr}\left(M_{q_D}U^\dagger\right) + \mathrm{h.c.} \right]
\end{align}
with the $SU(\nf)$ matrix $U \equiv \exp\left(2i\pid/\fpi \right)$, the dark pion decay constant $\fpi$, and the quark condensate $\mu^3_D$. 
In addition, the terms in Eq.~\eqref{eq:chiral_lagrangian} give rise to interactions between even numbers of dark pions. In the chirally broken phase the dark pions are the lightest mesons in the spectrum, and they can be stabilised by a suitable discrete or continuous symmetry.

In addition, there will also be heavier dark mesons in the spectrum, in particular vector mesons analogous to the SM $\rho$ mesons. Interactions between dark pions and dark vector mesons are introduced by using Massive Yang-Mills approach. The corresponding covariant derivative is given by~\cite{Sakurai}
\begin{align}
\label{eq:pion_covariant_derivative}
    D_\mu U = \partial_\mu U + i g [U, \rhod_\mu] \, ,
\end{align}
with the $\pid\pid\rhod$ coupling $g \approx m_\rhod/(\sqrt{2}\fpi)$ obtained from the KSRF relation~\cite{Kawarabayashi:1966kd,PhysRev.147.1071}. 
Hence, the chiral Lagrangian of dark pions and dark vector mesons, expanded up to terms with at most four dark pion fields, reads
\begin{align}
\mathcal{L}_\text{Ch} \supset &\mathrm{Tr}\left(D_\mu\pid D^\mu\pid\right) + m_\pid^2\mathrm{Tr}\left(\pi_D^2\right) + \tfrac{m_\pid^2}{3\fpi^2}\mathrm{Tr}\left(\pi_D^4\right) \nonumber\\  &-\tfrac{2}{3\fpi^2}\mathrm{Tr}\left(\pi_D^2 D_\mu\pid D^\mu\pid - \pid D_\mu\pid\pid D^\mu\pid\right) \; .
\end{align}
in agreement with the literature.
For $\nf \geq 3$, the Wess-Zumino-Witten (WZW) term induces an anomalous five-point interaction given by
\begin{align}
\mathcal{L}_\text{WZW} = \frac{2\nc}{15\pi^2\fpi^5} \epsilon^{\mu\nu\rho\sigma}\mathrm{Tr}\left(\pid\partial_\mu\pid\partial_\nu\pid\partial_\rho\pid\partial_\sigma\pid\right) \; .
\end{align}

The free parameters in the above chiral Lagrangian ($\mpi,\mrho,\fpi$) need to be calculated using non-perturbative methods. Along with $\nc, \nf$, a QCD-like strongly interacting theory with mass-degenerate quarks has two additional free parameters, one mass ratio and one mass scale~\cite{Albouy:2022cin}. Once these inputs are fixed, the dark meson spectrum can be computed using non-perturbative methods, e.g. lattice simulations. In the UV the two free parameters could be considered $M_{q_D}, M_{q_D}/\Lambda_D$, which are traded in for e.g. $\mpi, \mpi/f_{\pi_D}$ in the chirally broken phase.

In the following, we are particularly interested in the process $3\pid\to \pid\rhod$  (see Fig.~\ref{fig:3pi_to_pi_rho} for the corresponding Feynman diagrams) and its effect on the relic density of dark pions. For this process to be allowed in the non-relativistic limit (i.e.\ for negligible kinetic energy of the initial state particles), we require $\mrho/\mpi < 2$. Incidentally, this condition also implies that dark rho mesons cannot decay into pairs of dark pions. Nevertheless, dark rho mesons do in general decay into SM particles (for example through kinetic mixing with the SM photon~\cite{Holdom:1985ag,Babu:1997st}, see Refs.~\cite{Hochberg:2015vrg,Kribs:2018ilo,Katz:2020ywn, Bernreuther:2022jlj, Kulkarni:2022bvh} for explicit constructions). These (inverse) decays can efficiently transfer energy and entropy between the dark sector and the SM and keep the two sectors in thermal equilibrium. 

It will therefore be convenient to use $\mpi$ and the mass ratio $\mrho/\mpi$ as free parameters for our analysis. Using the results from Ref.~\cite{Maris:2005tt} and neglecting the effect of varying $\nc$ and $\nf$, we obtain an approximate relation between $\fpi$, $\mpi$ and $\mrho$:
\begin{align}
\label{eq:xi_mrhobympi_relation}
\xi \equiv \displaystyle\frac{\mpi}{\fpi} = 7.79 \, \displaystyle\frac{\mpi}{\mrho} + 0.57  \left(\displaystyle\frac{\mpi}{\mrho} \right)^2 \, ,
\end{align}
which is valid for $1 < \mrho/\mpi \lesssim 20$, where the lower limit on $\mrho/\mpi$ corresponds to $\xi \sim 8$. 
 
In principle, one could consider the entire range of above fit.  For large $\xi$, however, the validity of chiral perturbation theory becomes increasingly dubious and higher-order corrections become relevant~\cite{Hansen:2015yaa}. In the present work we therefore only consider dark rho meson masses larger than $1.45\,m_\pid$ (equivalently $\xi \lesssim 5.7$), corresponding to the range where the GMOR relation~\cite{Gell-Mann:1968hlm} is expected to be satisfied~\cite{Faber:2017alm, Engel:2014eea, Gattringer:2008vj, DelDebbio:2006cn}. This bound implies in particular that the forbidden annihilation process $\pid \pid \to \rhod \rhod$ studied in Ref.~\cite{Bernreuther:2019pfb} is not relevant for the calculation of the dark pion relic density. The case $m_\rhod > 2 m_\pid$, on the other hand, has previously been studied in Ref.~\cite{Choi:2018iit}. In this case, the process $3\pid \to 2\pid$ may receive a strong resonant enhancement from on-shell intermediate dark vector mesons. 

\smallskip

\paragraph*{Dark pion relic density.---}%
Let us now turn to the calculation of the dark pion relic density. The Boltzmann equation describes the evolution of the density of a particle species as it falls out of equilibrium. Integrating the Boltzmann equation yields the relic abundance of a stable particle given the contributing freeze-out processes. If $\rhod$ is in equilibrium with the SM bath throughout and $3\pid\to\pid\rhod$ annihilations dominate the depletion of dark pions, we can use the principle of detailed balance to write the Boltzmann equation as 
\begin{align}
\label{eq:boltzmanneq_2}
\dot{n}_\pid + 3H\,n_\pid = \langle \sigma v^2\rangle_{3\pid \to \pid \rhod} n_\pid ((n_\pid^\mathrm{eq})^2 - n_\pid^2) \; ,
\end{align}
where $n_\pid$ is the dark pion number density and $H$ denotes the Hubble rate.

In the non-relativistic limit, the thermally averaged cross section is given by
\begin{align}
    \langle\sigma v^2\rangle_{3\pid\to \pid \rhod} = & \frac{|\mathcal{M}|^2_{3\pid \to \pid\rhod}}{144\pi S_{\alpha} S_{\beta} m^3_\pid} \sqrt{4 - 5y + y^2} \, ,
\end{align}
with $S_\alpha=3!$ and $S_\beta=1$ denoting the number of permutations of identical particles in the initial and final state, respectively, and $y = m_\rhod^2 / (4 m_\pid^2)$.

Representative diagrams for the process are shown in Fig.~\ref{fig:3pi_to_pi_rho}.
Diagrams with an $s$-channel $\rho_D$ are velocity-suppressed and thus negligible in the final amplitude. The sum of all diagrams yields the total squared amplitude

\begin{widetext}
\begin{align}
|\mathcal{M}|^2_{3\pid \to \pid\rhod} = \frac{8 m_\pid^4(1-y)(4-y)(\frac{\Gamma_\mathrm{th}^2}{m_\pid^2} + 4y^2)(5\nf^4\frac{\Gamma_\mathrm{th}^2}{m_\pid^2} (13y+2)^2 + 32(2Ay^2+2By+C))}{3 f_\pid^6 (\frac{\Gamma_\mathrm{th}^2}{m_\pid^2} + 64) (2y+1)^2 (9 \frac{\Gamma_\mathrm{th}^2}{m_\pid^2} + 64 (1-y)^2))} \, ,
\label{eq:M2}
\end{align}
\end{widetext}

with 
\begin{eqnarray}
A &=& \frac{(821\nf^4-168\nf^2+36)}{\nf(\nf^2-1)^2} , \nonumber\\
B &=& \frac{(245\nf^4-114\nf^2+36)}{\nf(\nf^2-1)^2} ,\nonumber\\
C &=& \frac{(37\nf^4-30\nf^2+18)}{\nf(\nf^2-1)^2} ,
\end{eqnarray} 
and $\Gamma_\text{th}$ being the thermal width of dark pions discussed below. Details of this calculation can be found in App.~\ref{app:details}.

As $m_\rhod$ approaches $2 m_\pid$ from below, the internal dark pion can go on shell and the cross section grows rapidly. At first sight, this should lead to a divergence, since the dark pion is stable and therefore does not have a decay width. In a thermal plasma, however, dark pions do not have an infinite lifetime, due to interactions with other particles in the plasma. This leads to a thermal self-energy, the imaginary part of which can be interpreted as an effective decay width~\cite{Weldon:1983jn}.

We present the derivation of the dark pion thermal width $\Gamma_\text{th}$ in App.~\ref{app:thermal_width}. The main contribution is found to arise from the scattering of two dark pions. For $m_\rhod \approx 2 m_\pid$ and low temperatures, i.e.\ $x \equiv m_\pid / T \gg 1$, we find
\begin{align}
    \Gamma_\text{th} = & \frac{8\pi (\nf^2 - 1)}{x^2} e^{-x} m_\pid^3 \sigma_c \; , \label{eq:Gamma_th}
\end{align}
with  
\begin{align}
\label{eq:4pid_xsec}
    \sigma_{\mathrm{c}} &\approx \frac{1}{64\pi} \frac{3\nf^4-2\nf^2+6}{\nf^2(\nf^2-1)} \frac{m_\pi^2}{f_\pi^4} \nonumber\\
    &= \frac{3}{64\pi} \frac{m_\pi^2}{f_\pi^4} (1+\mathcal{O}(\nf^{-2}))
\end{align}
denoting the two-pion scattering cross section in the limit of vanishing initial velocities.

We emphasize that the thermal width becomes exponentially suppressed at small temperatures, as a result of the suppressed number density of dark pions that can participate in scattering. In the temperature range relevant for freeze-out ($x \approx 20$) the contribution from the dark pion width is therefore negligible unless $m_\rhod$ is extremely close to $2\,m_\pid$.

It is instructive to compare our result to the conventional freeze-out of SIMPs through the WZW anomaly. The corresponding Boltzmann equation reads
\begin{align}
\dot{n}_\pid + 3H\,n_\pid = \langle \sigma v^2\rangle_{3\pid \to 2\pid} n_\pid^2 (n_\pid^\mathrm{eq} - n_\pid) \; ,
\end{align}
where the thermally averaged cross section is given by~\cite{Kamada:2022zwb}
\begin{align}
\langle \sigma v^2 \rangle_{3\pid\to2\pid} = \frac{5\sqrt{5} \nc^2 \kappa_{3\pid\to2\pid} \xi^{10}}{4608\pi^5 m_\pid^5 x^2} \; ,
\end{align}
with
\begin{align}
    \kappa_{3\pid\to2\pid} = \frac{\nf (\nf^2-4)}{(\nf^2-1)^2} = \frac{1}{\nf} + \mathcal{O}(\nf^{-3}) \, .
\end{align}

The cross sections for the processes $3\pid\to \pid\rhod$ and $3\pid\to2\pid$ can both be written as
\begin{align}
\label{eq:alphaeff}
\langle\sigma v^2 \rangle \equiv \frac{\alpha^\mathrm{eff}}{m_\pid^5} \; .
\end{align}
To understand where the $3 \pid \to \pid \rhod$ process dominates over the well-known $3 \pid \to 2\pid$ SIMP process, it is useful to consider the ratio
\begin{align}
    \label{eq:R}
    R & \equiv \frac{\langle\sigma v^2\rangle_{3\pid\to\pid\rhod}}{\langle\sigma v^2\rangle_{3\pid\to2\pid}} = \frac{\alpha^\mathrm{eff}_{3\pid\to\pid\rhod}}{\alpha^\mathrm{eff}_{3\pid\to2\pid}} \nonumber \\
    & \approx \left(1800 \text{ -- } 8500\right) \times \frac{1}{\nc^2 \, \xi^4} \frac{x^2}{\sqrt{1-y}}\; ,\end{align}
where the range for the numerical factor has been obtained by varying $y$ in the range $0.6<y<1$ and $\nf$ between 3 and 6, noting that the process $3 \pid \to 2 \pid$ does not exist for $\nf = 2$~\cite{Bernreuther:2019pfb}. We find that the ratio $R$ is much larger than unity for all values that we consider. As expected, $R$ grows rapidly for $y \to 1$ and also grows with increasing $x$. This is because the process $3\pid \to 2 \pid$ proceeds via $d$-wave, whereas the process $3\pid \to \pid \rhod$ proceeds via $s$-wave. We conclude that the latter process will completely dominate dark pion freeze-out for $m_\rhod < 2 m_\pid$. As shown in App.~\ref{app:visible}, also direct annihilations of the dark pions into SM fermions via an off-shell dark rho meson are negligible as long as the SM coupling of the dark rho meson is sufficiently small.

\smallskip 
\paragraph*{Results and discussion.---}

\begin{figure}
    \centering
    \includegraphics[width=1.0\linewidth]{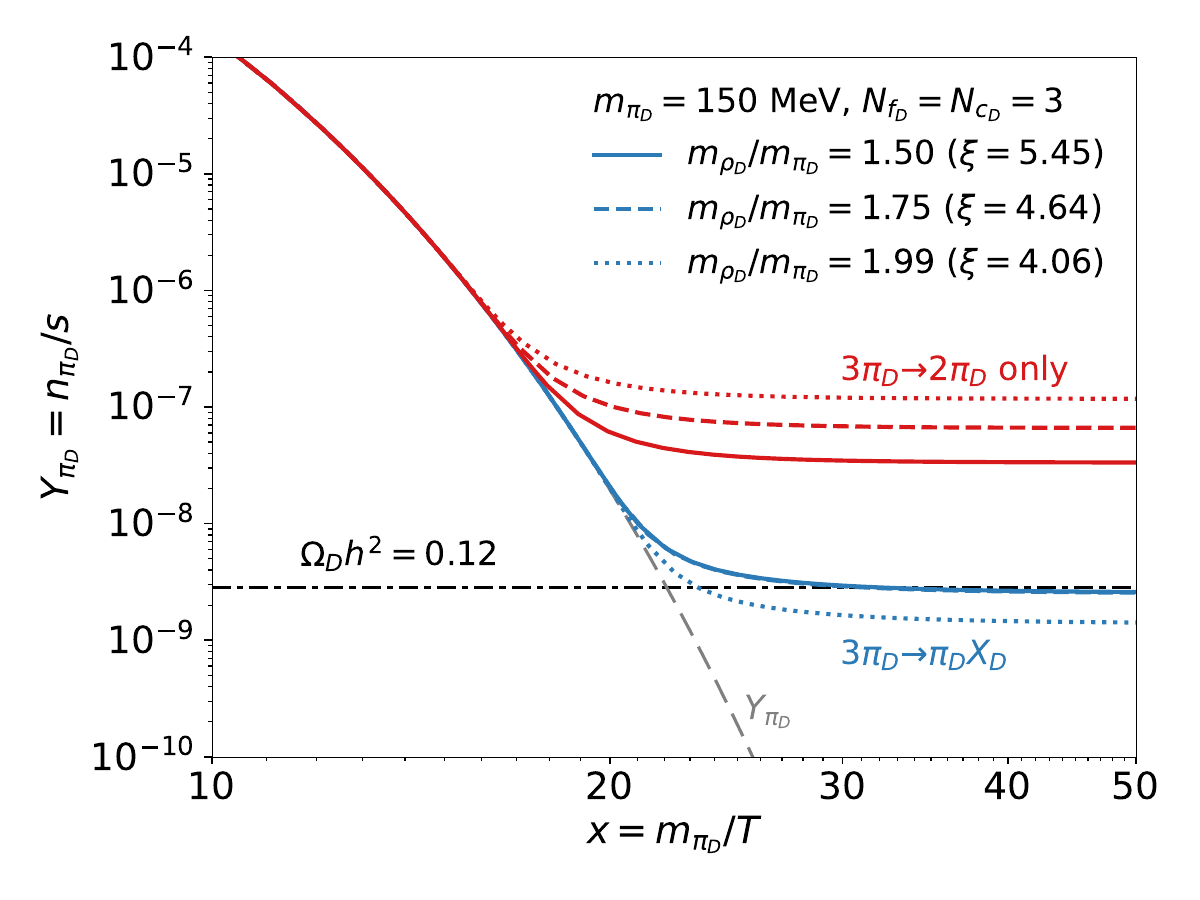}
    \caption{Solutions of the Boltzmann equation for the (dimensionless) dark pion number density as a function of inverse temperature. For the red lines only the process $3\pid\to2\pid$ is included, whereas the blue lines include $3\pid\to\pid X_D$ with $X_D = \pid,\rhod$. The value of $Y_\pid$ that corresponds to the observed DM relic abundance is indicated by the dot-dashed black line, while the equilibrium value of $Y_\pid$ is represented by the dashed grey line.}
    \label{fig:boltzmannSolution}
\end{figure}

It is convenient to express the Boltzmann equation in terms of the dimensionless quantity $Y_\pid=n_\pid/s$, where $s = 2\pi^2/45 \, g_\star T^3$ is the total entropy density. The Boltzmann equation~\eqref{eq:boltzmanneq_2} then takes the form
\begin{align}
    \label{eq:boltzmanneq_y}
    \frac{dY}{dx} = \frac{s^2}{\tilde{H}x}\langle\sigma v^2\rangle \: Y \: (Y_\mathrm{eq}^2-Y^2) \; ,
\end{align}
where following Ref.~\cite{bringmann2022darksusy} we have introduced the modified Hubble rate $\tilde{H} = (\frac{8\pi^3}{90}g)^{1/2} \frac{T^2}{M_\mathrm{Pl}} [1+\frac{1}{3}\frac{\text{d(ln }g_\star)}{\text{d(ln T)}}]^{-1}$, with $g$ ($g_\star$) denoting the energy (entropy) degrees of freedom taken from Ref.~\cite{Drees_2015}. We solve the Boltzmann equation in log-space.

In Fig.~\ref{fig:boltzmannSolution} we show the solution of the Boltzmann equation for $m_\pid = 150\,\mathrm{MeV}$, $\nf = \nc = 3$ and different values of $m_\rhod / m_\pid$. As expected, the dimensionless dark pion number density $Y_\pid$ follows the equilibrium value $Y_\pid^\text{eq} = (\nf^2 - 1) m_\pid^3 (x / (2 \pi))^{3/2} \exp (-x)$ until it freezes out at $x_f \approx 20$ and becomes constant. For the chosen parameters, the observed DM relic abundance $\Omega_D h^2 = 0.12$ is approximately reproduced. For comparison, we also show the evolution of $Y_\pid$ obtained when only the process $3\pid \to 2\pid$ is considered~\cite{Note1}. For the same masses, the predicted relic abundance is too large by approximately an order of magnitude.

We can understand this result analytically, by writing Eq.~\eqref{eq:boltzmanneq_y} as
\begin{align}
\label{eq:boltzmanneq_y2}
\frac{dY}{dx} =  \frac{\lambda_{3\to 2}}{x^5}\: Y \: (Y_\mathrm{eq}^2-Y^2) \; ,
\end{align}
with
\begin{align}
    \lambda_{3\to 2} = \frac{2\sqrt{5}}{675}  \pi^{5/2} \, g^{3/2} \, M_\mathrm{Pl} \, m_\pid^4 \, \langle\sigma v^2 \rangle \, ,
\end{align}
where $g=g_*=10.75$ in the temperature range of interest and we can drop the derivative in the expression for $\tilde{H}$.

Typically, $ Y_\mathrm{eq} \ll Y$ during freeze-out and the corresponding term in the Boltzmann equation can be neglected. Treating $\lambda_{3 \to 2}$ as a constant during freeze-out by setting $\lambda_{3 \to 2} \equiv \lambda_{3 \to 2}(x=x_f)$, Eq.~\eqref{eq:boltzmanneq_y2} can be approximately solved analytically, which yields the asymptotic solution
\begin{align}
Y_\infty \approx \sqrt{2} \frac{x_f^2}{\sqrt{\lambda_{3\to 2}}} \, .
\end{align}
Hence, the DM relic abundance scales with the DM mass and effective coupling (as defined in Eq.~\eqref{eq:alphaeff}) as $\Omega_D h^2 \sim m_\pid^{3/2} / \sqrt{\alpha_\mathrm{eff}}$.
Therefore, when $3\pid\to\pid\rhod$ annihilations dominate over $3\pid\to2\pid$, the preferred dark pion mass scale increases as $m_\pid \sim R^{1/3}$ relative to the usual expectation for SIMPs annihilating via the WZW term. For typical values of $\xi$, $m_\rhod / m_\pid$ and $\nc$ this corresponds to a factor of 2--3, and even more for $m_\rhod \to 2 m_\pid$, see Eq.~\eqref{eq:R}.

\begin{figure}
    \centering
    \includegraphics[width=1.0\linewidth]{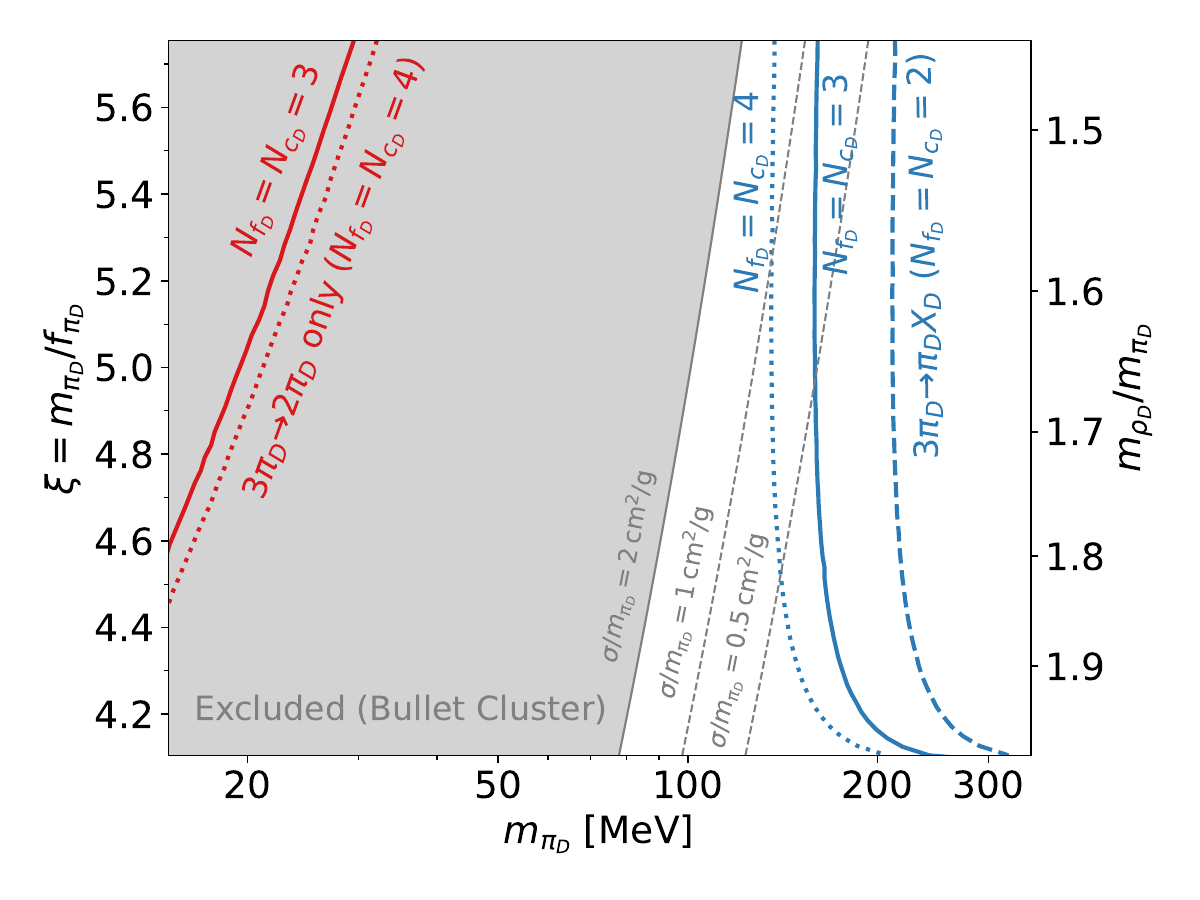}
    \caption{Combinations of $m_\pid$ and $\xi = m_\pid / f_\pid$ (or $m_\rhod / m_\pid$ via Eq.~\eqref{eq:xi_mrhobympi_relation}) that yield $\Omega_D h^2 = 0.12$ in agreement with observations. For the red lines only the process $3\pid\to2\pid$ is included, whereas for the blue lines $3\pid\to\rhod\pid$ is also taken into consideration.
    Note that the process $3\pid \to 2\pid$ does not exist for $\nf = 2$. The grey shaded region is excluded by the Bullet Cluster bound on the DM self-interaction cross section (evaluated for $\nf \gg 1$).}
    \label{fig:3}
\end{figure}

We explicitly confirm this expectation in Fig.~\ref{fig:3}, where we plot the combinations of $m_\pid$ and $\xi$ (or equivalently $m_\rhod / m_\pid$ as related to $\xi$ by Eq.~\eqref{eq:xi_mrhobympi_relation}) that yield $\Omega_\mathrm{D} h^2 = 0.12$ via a numerical solution of the Boltzmann equation for different choices of $\nf$ and $\nc$. As before, we show for comparison the result when considering $3\pid \to 2 \pid$ only, which is robustly excluded by the Bullet Cluster constraint on DM self-interactions, given by~\cite{Robertson:2016xjh,Wittman:2017gxn}
\begin{align}
    \frac{\sigma_c}{m_\pid} \lesssim 2~\mathrm{cm^2 / g} \; ,
\end{align}
with $\sigma_c$ as in Eq.~\eqref{eq:4pid_xsec}.
Including the dark rho mesons in the final state, on the other hand, increases the dark pion mass scale sufficiently to evade these constraints. Due to the interplay between $\xi$ and $m_\rhod / m_\pid$ in Eq.~\ref{eq:M2}, we find that the preferred value of $m_\pid$ is largely independent of these parameters and can be approximately written as $m_\pid \approx 330 \, \mathrm{MeV} / \nf^{2/3}$. 

\smallskip

In conclusion, we have shown that for strongly interacting dark sectors with $m_\rhod / m_\pid < 2$, the dominant process that changes the number density of dark pions in the early universe is $3\pid \to \pid \rhod$. We emphasize that this process does not depend on any non-perturbative parameters other than $m_\pid$, $f_\pid$ and $m_\rhod$ and~-- in contrast to the conventionally studied process $3\pid \to 2\pid$~-- does not rely on the WZW anomaly, i.e.\ it also exists for theories with only two light flavours. In contrast to the recently proposed Co-SIMP mechanism~\cite{Smirnov:2020zwf,Parikh:2023qtk}, the process that we consider requires no interactions between the dark sector and SM particles beyond those needed for thermalisation (which only places a very weak lower bound on the decay width of the dark rho mesons). As a result, we obtain a theoretically clean and robust prediction for the dark pion mass that reproduces the observed DM relic abundance. We have specifically considered the range $1.45 \leq m_\rhod / m_\pid < 2$ corresponding to $4 \lesssim m_\pid / f_\pid \lesssim 6$, within the validity of chiral perturbation theory. We find typical values of $m_\pid$ around $100\,\mathrm{MeV}$ reproduce the measured DM relic abundance, with only mild dependence on the other parameters. Crucially, these dark pion masses satisfy the Bullet Cluster constraint on DM self-interactions -- unlike the masses favoured by freeze-out via $3\pid\to2\pid$ -- and therefore provide an important benchmark scenario for further exploration. We emphasize that the required mass spectrum implies that the dark rho mesons can only decay into SM particles, leading to exciting signatures at laboratory experiments.

\acknowledgments

We thank Pieter Braat, Gilberto Colangelo, Maximilian Detering, Gernot Eichmann, Mads Frandsen, Bror Hjemgaard, Martin Hoferichter, Martin Napetschnig,  Marieke Postma and Fabian Zierler for helpful discussions. NH thanks CP$^3$-Origins for hospitality.
Fermilab is administered by Fermi Research Alliance, LLC under Contract No.\ DE-AC02 07CH11359 with the U.S.\ Department of Energy, Office of Science, Office of High Energy Physics. NH and FK are supported by the Deutsche Forschungsgemeinschaft (DFG) through the Emmy Noether Grant No.\ KA 4662/1-2 and grant 396021762~--~TRR~257. SK is supported by FWF research group grant FG1. 

\providecommand{\bysame}{\leavevmode\hbox to3em{\hrulefill}\thinspace}

\appendix

\begin{widetext}

\section{Matrix elements}
\label{app:details}

Here we present the matrix elements corresponding to the $3\pid \to \pid \rhod$ annihilation diagrams shown in Fig.~\ref{fig:3pi_to_pi_rho}.

The matrix element for the contact-interaction contribution is given by
\begin{align}
    i \mathcal{M} = \frac{2 i g}{3 f_\pid} \epsilon^\star_\mu(p_e) (p_d^\mu f_{abcde} + p_a^\mu f_{bcdae} + p_b^\mu f_{cdabe} + p_c^\mu f_{dabce}) \; ,
\end{align}
where $p_a$, $p_b$, $p_c$, $p_d$ denote the dark pion momenta, $p_e$ is the dark rho momentum, and all momenta are defined as incoming. $\epsilon^\star_\mu(p_e)$ denotes the polarization vector of the dark rho. 
The group generators of $SU(\nf)$ enter via
\begin{align}
    f_{abcde} \equiv \, &T_{ab c d e} + T_{a c b d e} + T_{bacde} + 
 T_{b c a d e} + T_{c a b d e} + T_{c b a d e} \; ,
\end{align}
with
\begin{align}
    T_{a b c d e} \equiv & 2\,\mathrm{Tr}[\, T^a T^b T^e T^c T^d] - 2\,\mathrm{Tr}[T^a T^e T^b T^c T^d] - 
  \mathrm{Tr}[T^a T^b T^c T^e T^d] + \mathrm{Tr}[T^a T^b T^e T^c T^d] \nonumber\\ &- \mathrm{Tr}[T^a T^b T^d T^c T^e] + 
  \mathrm{Tr}[T^a T^b T^d T^e T^c]] \; .
\end{align}

The matrix elements for the contributions with one 4$\pid$ vertex and one $\pid\pid\rhod$ vertex connected by an internal pion propagator are given by
\begin{align}
    i \mathcal{M} = &\frac{-2 i g}{3 f_\pid^2 ((p_a+p_e)^2-m_\pid^2 + i(E_a+E_e)\Gamma_\mathrm{th})} \epsilon^\star_\mu(p_e) \nonumber\\
    & \times \left[p_a^\mu (\mathrm{Tr}[T^aT^fT^e]-\mathrm{Tr}[T^aT^eT^f]) - (p_a^\mu+p_e^\mu)(\mathrm{Tr}[T^fT^aT^e]-\mathrm{Tr}[T^fT^eT^a]) \right] \nonumber\\ & \times \left[ \left(-4 p_b \cdot p_d - 4 (p_a+p_e) \cdot p_c + 2 p_c \cdot p_d + 2 p_b \cdot p_c + 
   2 (p_a+p_e) \cdot p_b + 2 p_d \cdot (p_a+p_e) - 4 m_\pid^2\right) \mathrm{Tr}[T^f T^b T^c T^d] \right. \nonumber\\ & \quad + \left(-4 p_b \cdot p_c - 4 (p_a+p_e) \cdot p_d + 2 p_d \cdot  p_c + 2 p_b \cdot p_d + 
   2 (p_a+p_e) \cdot p_b + 2 p_c \cdot (p_a+p_e) - 4 m_\pid^2\right) \mathrm{Tr}[
  T^f T^b T^d T^c]\nonumber\\ & \quad + \left(-4 p_c \cdot p_d - 4 (p_a+p_e) \cdot p_b + 2 p_b \cdot p_d + 2 p_c \cdot p_b + 
   2 (p_a+p_e) \cdot p_c + 2 p_d \cdot (p_a+p_e) - 4 m_\pid^2\right) \mathrm{Tr}[
  T^f T^c T^b T^d]\nonumber\\ & \quad + \left(-4 p_c \cdot p_b - 4 (p_a+p_e) \cdot p_d + 2 p_d \cdot p_b + 2 p_c \cdot p_d + 
   2 (p_a+p_e) \cdot p_c + 2 p_b \cdot (p_a+p_e) - 4 m_\pid^2\right) \mathrm{Tr}[
  T^f T^c T^d T^b] \nonumber\\ & \quad + \left(-4 p_d \cdot p_c - 4 (p_a+p_e) \cdot p_b + 2 p_b \cdot p_c + 2 p_d \cdot p_b + 
   2 (p_a+p_e) \cdot p_d + 2 p_c \cdot (p_a+p_e) - 4 m_\pid^2\right) \mathrm{Tr}[
  T^f T^d T^b T^c]\nonumber\\ & \quad + \left. \left(-4 p_d \cdot p_b - 4 (p_a+p_e) \cdot p_c + 2 p_c \cdot p_b + 2 p_d \cdot p_c + 
   2 (p_a+p_e) \cdot p_d + 2 p_b \cdot (p_a+p_e) - 4 m_\pid^2\right) \mathrm{Tr}[
  T^f T^d T^c T^b] \right] \, ,
\end{align}
and corresponding expressions with $(a \leftrightarrow b)$, $(a \leftrightarrow c)$, and $(a \leftrightarrow d)$, with the thermal dark pion width $\Gamma_\mathrm{th}$ as discussed above.

The diagrams with one $\pid\pid\rhod$ and one $\pid\pid\rhod\rhod$ vertex connected by an internal rho propagator correspond to the matrix element

\begin{align}
    i \mathcal{M} = &\frac{-2i g^3}{(p_a+p_d)^2 - m_\rhod^2 + i(E_a+E_d)\Gamma_\rhod} \epsilon^\star_\mu(p_e) \nonumber\\  & \times (
     p_a^\mu (\mathrm{Tr}[T^aT^dT^f]] - \mathrm{Tr}[T^aT^fT^d]) + 
   p_d^\mu (\mathrm{Tr}[T^dT^aT^f] - \mathrm{Tr}[T^dT^fT^a]))\nonumber\\ &\times (\mathrm{Tr}[T^cT^eT^bT^f] - \mathrm{Tr}[T^cT^eT^fT^b) + \mathrm{Tr}[T^cT^fT^bT^e] - \mathrm{Tr}[T^cT^fT^eT^b] \nonumber\\ &\quad + \mathrm{Tr}[T^bT^eT^cT^f] - \mathrm{Tr}[T^bT^eT^fT^c] + \mathrm{Tr}[T^bT^fT^cT^e] - \mathrm{Tr}[T^bT^fT^eT^c] \nonumber\\ &\quad - \mathrm{Tr}[T^eT^cT^bT^f] + \mathrm{Tr}[T^eT^cT^fT^b] - \mathrm{Tr}[T^eT^bT^cT^f] + \mathrm{Tr}[T^eT^bT^fT^c] \nonumber\\ &\quad - \mathrm{Tr}[T^fT^cT^bT^e] + \mathrm{Tr}[T^fT^cT^eT^b] - \mathrm{Tr}[T^fT^bT^cT^e] + \mathrm{Tr}[T^fT^bT^eT^c]) \, ,
\end{align}

and analogous expressions with $(a \leftrightarrow b)$, and $(a \leftrightarrow c)$.

The contributions with both an internal pion propagator and an internal rho propagator yield

\begin{align}
    i \mathcal{M} = &-\frac{8 i g^3}{((p_a+p_e)^2-m_\pid^2 + i(E_a+E_e)\Gamma_\pid)((p_c+p_d)^2-m_\rhod^2 + i(E_c+E_d)\Gamma_\rhod)} \epsilon^\star_\mu(p_e) \nonumber\\ & \quad \times (p_a^\mu (\mathrm{Tr}[T^aT^fT^e]- \mathrm{Tr}[T^aT^eT^f]) - (p_a+p_e)^\mu (\mathrm{Tr}[T^fT^aT^e]- \mathrm{Tr}[T^fT^eT^a]))  \nonumber\\ & \quad \times (p_{d,\nu} (\mathrm{Tr}[T^dT^cT^h]- \mathrm{Tr}[T^dT^hT^c]) + p_{c,\nu} (\mathrm{Tr}[T^cT^dT^h]- \mathrm{Tr}[T^cT^hT^d])) \nonumber\\ & \quad \times (p_b^\nu (\mathrm{Tr}[T^bT^fT^h]- \mathrm{Tr}[T^bT^hT^f]) + (p_a+p_e)^\nu (\mathrm{Tr}[T^fT^bT^h]- \mathrm{Tr}[T^fT^hT^b])) \, ,
\end{align}

as well as analogous matrix elements with $(a \leftrightarrow b)$, $(a \leftrightarrow c)$, $(b \leftrightarrow c)$, $(abc \leftrightarrow cab)$, and $(abc \leftrightarrow bca)$. 

To calculate squared matrix elements we use the \textsc{FeynCalc}~\cite{Mertig:1990an, Shtabovenko:2016sxi, Shtabovenko:2020gxv} Mathematica package.
\end{widetext}

\section{Thermal width of dark pions}
\label{app:thermal_width}

At zero temperature, the dark pions are stable, and hence their propagator is given by
\begin{equation}
    D = \frac{1}{q^2 - m_\pid^2} \; ,
\end{equation}
which diverges for $q^2 = m_\pid^2$. At finite temperature, this expression is modified to
\begin{equation}
    D = \frac{1}{q^2 - m_\pid^2 - \Pi(q^2)} \; ,
\end{equation}
where $\Pi(q^2)$ denotes the self-energy of the dark pion. The real part of the self-energy amounts to a thermal mass, which shifts the position of the pole of the propagator. For temperatures small compared to the dark pion mass (as relevant for freeze-out calculations), this effect is negligible. The imaginary part of the self-energy, on the other hand, provides a thermal width, which regulates the divergence of the propagator at finite temperatures. We can then write the propagator in the usual Breit-Wigner form
\begin{equation}
    D = \frac{1}{q^2 - m_\pid^2 + i E_\pid \Gamma_\text{th}}
\end{equation}
with
\begin{equation}
    \Gamma_\text{th} = - \frac{1}{E_\pid} \text{Im} \, \Pi \; .
\end{equation}
Note that the thermal width depends on the energy $E_\pid$ of the pion relative to the plasma (i.e.\ in the cosmic rest frame).

The thermal width can be interpreted as the rate at which the phase space distribution approaches equilibrium~\cite{Weldon:1983jn}. In other words, $\Gamma_\text{th}$ is given by the interaction rate of the dark pion with the plasma. 
Schematically,
\begin{align}
    \text{Im} \, \Pi = - \frac{1}{2 (1 + f_\pid(E_\pid))} \sum_{X,Y} & \int \mathrm{d} \Pi_X \mathrm{d} \Pi_Y |\mathcal{M}_{\pi + X \to Y}|^2 \nonumber \\ & \times (2\pi)^4 \delta^4\left(p_\pi + p_X - p_Y\right)
\end{align}
where the sum is over all kinematically allowed processes and $\mathrm{d} \Pi_{X,Y}$ denote the phase space volume multiplied with the appropriate distribution function, i.e.
\begin{equation}
    \mathrm{d}\Pi_{X,i} = \frac{\mathrm{d}^3 p_i}{2 E_i (2\pi)^3} f(E_i)
\end{equation}
for a particle in the initial state and
\begin{equation}
    \mathrm{d}\Pi_{Y,i} = \frac{\mathrm{d}^3 p_i}{2 E_i (2\pi)^3} (1 \pm f(E_i))
\end{equation}
for a boson (positive sign) or fermion (negative sign) in the final state.

Since the pion is the lightest particle of the theory, there are no $1 \to n$ processes with real particles in the final state, and hence the leading diagrams contributing to the thermal width are $2 \to 2$ processes, i.e.\ $\pid \pid \to \pid \pid$ and $\pid \rhod \to \pid \rhod$. The latter is suppressed relative to the former by the ratio of phase-space densities $f_\rhod(E_i) / f_\pid(E_i)$ and is therefore negligible. The thermal width is then given by~\cite{Note2}\begin{equation}
 \Gamma_\text{th}(E_1) = \int \frac{\mathrm{d}^3 p_2}{(2\pi)^3} f_\pid(E_2) \sigma(\pid_1 \pid_2 \to \pid_3 \pid_4) v_\text{rel} \; ,
\end{equation}
where we assume that the pions are non-relativistic, such that $f_\pid(E) \ll 1$ and hence $1 + f_\pid(E) \approx 1$. In this case we recover the standard phase space integral for a $2 \to 2$ cross section at zero temperature. 

To obtain the thermal width, we need to calculate the cross section of $\pid\pid \to \pid\pid$ scattering. The matrix element obtains contributions from the four-pion-contact interaction, and from virtual $\rhod$ exchange in $s$-, $t$- and $u$-channel diagrams. The the latter contributions are momentum-suppressed and hence subdominant in the non-relativistic limit~\cite{Note3}. We can therefore approximate the total cross section by the cross section for contact interaction
\begin{align}
    \sigma_{\mathrm{c}} = \frac{1}{2\pi s} \frac{1}{f_\pid^4} \Bigg( & \frac{3\nf^4-2\nf^2+6}{8\nf^2(\nf^2-1)} m_\pid^4 \Bigg. \nonumber \\ & \Bigg. + \frac{\nf^2}{\nf^2-1}\left(m_\pid^2p^2 + \frac{5}{6} p^4\right) \Bigg) \; ,
\end{align}
with $p$ denoting the momentum of each of the incoming pions in the center-of-mass frame. In the non-relativistic limit we have
\begin{equation}
    \sigma_{\mathrm{c}} \approx \frac{1}{64\pi} \frac{3\nf^4-2\nf^2+6}{\nf^2(\nf^2-1)} \frac{m_\pid^2}{f_\pid^4} \; .
\end{equation}

For non-relativistic dark pions, the phase space distribution is given by
\begin{equation}
    f_\pid(E) =  \left(\nf^2 - 1\right) e^{-\left(x+\tfrac{1}{2} x v^2\right)} \; ,
\end{equation}
where $x = m_\pid / T$. 
We therefore need to calculate the integral
\begin{equation}
I =  \int \mathrm{d}^3 v_2 \, \sigma(v_\text{rel}) v_\text{rel} e^{-x v_2^2 / 2} \; .
\end{equation}
To do so, we note that $\mathbf{v}_2 = \mathbf{v}_\text{rel} + \mathbf{v}_1$ and obtain
\begin{align}
I & = \frac{\pi}{x^2}  \sigma_\text{c} \left[4 e^{-w_1^2/2} + \frac{
   2 \sqrt{2 \pi} (1 + w_1^2) \text{erf}(w_1/\sqrt{2})}{w_1} \right]  \nonumber\\ & \approx \frac{8\pi}{x^2}  \sigma_\text{c} \; , \label{eq:approx}
\end{align}
where we have defined $w_1 = v_1 \sqrt{x}$ and the second line holds for $w_1 \ll 1$.

When calculating the cross section for $3\pid \to \pid \rhod$, we set the velocities of all initial state particles to zero. The velocity $v$ of the virtual pion is therefore fixed by energy and momentum conservation:
\begin{equation}
 v_1 = \frac{2 \sqrt{4 - 5 y + y^2}}{3} \; ,
\end{equation}
where $y = m_\rhod^2 / (4 m_\pid^2)$~\cite{Note4}.

Putting everything together, we then obtain Eq.~(11) from the main text. It is worth noting that, due to the exponential suppression of pions in the plasma, the thermal width is extremely small and rapidly decreases with temperature. Explicit calculations show that for any of the points in parameter space that we consider, the thermal width contributes less than a percent of the total cross section, and around the time of freeze-out, at $x=x_f$, the contribution is negligible.

\section{Interactions between the dark and visible sector}
\label{app:visible}

The number-changing processes in the dark sector convert rest mass to kinetic energy. In the absence of interactions between the dark and the visible sector, this would modify the temperature evolution of the dark sector and lead to a dark sector that is hotter than the SM thermal bath. As a result, the relic density calculation would be modified and larger couplings (or smaller masses) would be needed to avoid overproducting DM.

For our calculations to be valid, the $\rhod$ meson is hence assumed to couple to SM particles at a level sufficient to ensure thermal equilibrium. In the mass range of interest, the most relevant coupling is the one of the $\rhod$ meson to electrons and positrons, which we can in general write as
\begin{equation}
    \mathcal{L}_\text{mix} = g_\ell \bar{e} \gamma_\mu e \rhod^\mu \; ,
\end{equation}
such that
\begin{equation}
    \Gamma_\rhod = \frac{m_\rhod g_\ell^2}{12 \pi} 
\end{equation}
for $m_\rhod \gg m_e$.

The origin of the coupling $g_\ell$ depends on the details of how the two sectors interact. For example, we can assume that the dark quarks carry charge $\pm 1$ under a new $U(1)'$ gauge group with gauge coupling $e_\mathrm{D}$. If the corresponding gauge boson $Z'$ has mass $m_{Z'}$ and kinetic mixing $\epsilon$ with the SM photon, one finds~\cite{Bernreuther:2019pfb}
\begin{equation}
    g_\ell = \frac{m_\rhod^2}{m_{Z'}^2} \frac{2 e e_\mathrm{D} \kappa}{g} \; .
\end{equation}

The Boltzmann equation for decays of the $\rhod$ mesons is given by
\begin{equation}
\frac{\mathrm{d} Y}{\mathrm{d} x} = - \frac{\Gamma_\rhod}{H x} \left( \frac{Y}{Y^\text{eq}} - 1 \right) \; , 
\end{equation}
where $Y = n_\rhod / s$ and $Y^\text{eq}$ is the corresponding equilibrium distribution.
Defining $y = Y / Y^\text{eq}$ and using $Y^\text{eq} \propto e^{-x} x^{3/2}$, this equation becomes
\begin{equation}
 x \frac{\mathrm{d}y}{\mathrm{d} x} = - \frac{\Gamma_\rhod}{H} \left(y - 1\right) + y \left(x - \frac{3}{2}\right) \; .
\end{equation}
In order for this equation to restore equilibrium ($y \approx 1$) it is not sufficient to require $\Gamma_\rhod/H \gg 1$, but we need to require $\Gamma_\rhod/H \gg x$, such that the second term on the right-hand side is negligible. This is different from the case of annihilations, where the first term is proportional to $y^2 - 1$ and therefore always dominates over the second term for large deviations from equilibrium.

For $m_\pid \sim 100 \, \mathrm{MeV}$, $m_\rhod \sim 2 m_\pid$ and $x \sim 20$ the requirement $\Gamma_\rhod / H \gg x$ implies $g_\ell \gtrsim 10^{-9}$. This lower bound can be compared to the upper bound on $g_\ell$ from the requirement that the direct annihilation process $\pi^+ \pi^- \to e^+ e^-$ does not modify the relic density calculation. We find that the corresponding cross section is given by~\cite{Braat:2023fhn}
\begin{align}
    & \langle \sigma v \rangle_{\pid^+ \pid^- \to e^+ e^-} \nonumber =  \\ & \frac{g_\ell^2 \, \alpha_D}{3\sqrt{\pi}}\frac{\nf}{\nf^2-1} \frac{x^{3/2}e^{2x}}{m_\pid^2}  \int_1^\infty d\Tilde{s}  \frac{\Tilde{s}^{3/4} (\Tilde{s}-1)^{3/2} e^{-2x\sqrt{\Tilde{s}}}}{(\Tilde{s}-\frac{m_\rhod^2}{4m_\pid^2})^2} ,
\end{align}
where $\Tilde{s}=s / (4m_\pid^2)$. This cross section is found to give a negligible contribution to the freeze-out calculation for $g_\ell \lesssim 3 \times 10^{-5}$. The cross section for $\pid \pid \to \pid e^+ e^-$ via a virtual $\rhod$ meson is even smaller in the non-relativisitc limit, where the kinetic energy of the initial state is insufficient to produce the $\rhod$ meson on-shell.

We conclude that there is a wide region of parameter space, where the assumptions made in our analysis are satisfied, i.e.\ the decays of the dark rho meson maintain thermal equilibrium between the dark and visible sectors, but direct annihilations of dark pions into SM final states are negligible.

\end{document}